\documentclass[10pt]{article}
\usepackage{sao2}
\usepackage{psfig,epsf}

\issue{2011, 55, 392--399}

\begin{document}
\markboth{M.L.~Khabibullina, O.V.~Verkhodanov, M.~Singh, A.~Pirya, S.~Nandi, N.V.~Verkhodanova}
  {A SECOND SET OF RATAN-600 OBSERVATIONS OF GIANT RADIO GALAXIES}
\title{A Second Set of RATAN-600 Observations of Giant Radio Galaxies}
\author{M.L.~Khabibullina\inst{a},
O.V.~Verkhodanov\inst{a},
M.~Singh\inst{b},
A.~Pirya\inst{b},
S.~Nandi\inst{b},
N.V.~Verkhodanova\inst{a}
}
\institute{
$^a$\saoname; \\ $^b$\ARIES}

\date{September 2, 2010}{November 20, 2010}
\maketitle

\begin{abstract}
Results of RATAN-600 centimeter-wavelength flux-density measurements of the
extended components in five giant radio galaxies are reported. The spectra
of the components of these radio galaxies have been constructed using the
data of the WENSS, NVSS, and GB6 surveys together with new RATAN-600 data.
Spectral indices in the studied frequency range have been calculated.\\

PACS:  98.54.Gr, 98.62.Ai, 98.70.Dk, 98.70.Vc
\end{abstract}
\maketitle

\section{INTRODUCTION}
In 2008, we initiated a program of observations of giant radio galaxies
(GRGs) on the RATAN-600 radio telescope. In December 2008, we conducted
a first set of observations and obtained integrated spectra for the
components in seven radio galaxies: GRG\,0139+3957, GRG\,0912+3510,
GRG\,1032+2759, GRG\,1343+3758, GRG\,1400+3017, GRG\,1453+3308, GRG\,1552+2005,
and GRG\,1738+3733. These observations \cite{kh_a} showed that the radio spectral
indices of the components of these radio galaxies differ considerably.
The form of the spectra also differs, from very steep, as in GRG\,0139+3957,
to fairly flat, as in the components of GRG\,1738+3733.

Radio observations of GRGs can help elucidate the origin of objects with such
 giant sizes, which is not yet entirely clear. According to the generally
accepted definition, GRGs are radio sources with linear dimensions larger than
1 Mpc. They are the largest objects in the visible Universe, and it is possible
that they played a special role in the formation of the large-scale structure.
The large sizes of GRGs also suggest that these sources must be at a late stage
in their evolution. Models of radio sources \cite{kaiser_a,blundell} predict a change in the radio
luminosity and linear dimensions of powerful radio sources with time. According
to these models, GRGs should be very old objects (with ages >108 years) and should
presumably be in a lower density medium compared to smaller sources with comparable
radio luminosities \cite{kaiser_b}. Their analysis of
radio and optical data (SDSS, APM) for radio galaxies and quasars led Komberg
and Pashchenko \cite{komberg} to conclude that, apart from environment effects, the sizes
of giant radio sources could be explained by the presence of a population of
long-lived, radio-loud active galactic nuclei, which, in turn, can evolve to
GRGs. Multifrequency observations \cite{mack} showed that the spectral ages of GRGs are
greater than is expected from evolutionary models. As was noted in \cite{jamrozy_a}, such
radio galaxies could influence the formation of galaxies, since the pressure of
the gas outflowing from a radio source could compress cold gas clouds and
initiate the development of stars, on the one hand, and stop the formation of a
galaxy, on the other hand. Note that several teams \cite{schoenmakers_a}-\cite{nandi}
are investigating the
origin of GRGs and physical processes in these radio sources. However, thus far,
there is no definite solution to this problem.

Our analysis of the spectra of radio galaxies in \cite{kh_a} led us to the conclusion that
the center-to-edge
variation of the spectral indices for GRGs is due to a change in the particle
energy in the components (this was also noted earlier in \cite{schoenmakers_b}. This change is due
to pressure variations in the surrounding streaming gas; i.e., to a change in the
environment that depends on distance from the host galaxy. However, general
conclusions will be more significant in an integrated approach to the GRG population
as a whole. An important stage in studying the origin of the large sizes of GRGs
is a comparative analysis of the properties of "ordinary" radio galaxies, also carried
out in the framework of this project \cite{kh_b}-\cite{ver_a}. Earlier, Soboleva
\cite{soboleva} observed arcminute-size radio galaxies at centimeter and decimeter
wavelengths on the RATAN-600, and found that their morphological structures
have virtually identical spectral indices. Therefore, studies of the
remaining objects of the GRG sample will enable us to supplement available
information on the radio spectra of this population.

We also noted that the millimeter-wavelength flux densities of the studied GRG
components estimated from extrapolated spectra are above 0.6 mJy \cite{kh_a}. This
could be relevant for estimates of the contributions of various effects to
anisotropy of the cosmic microwave background (CMB) on scales of galaxy
clusters. Thus, knowledge of the spectral indices, shapes, and orientations
of GRGs will facilitate taking into account features of regions with sizes of
up to 10' on maps of the CMB anisotropy \cite{ver_b}.

In the current paper, we continue our RATAN-600 study of this GRG sample and
present new observations of these radio galaxies at centimeter and decimeter
wavelengths.

\section{RATAN-600 DATA}
\subsection{RATAN-600 Observations}

The observations of a new subsample of GRGs were conducted on the Southern
sector of the RATAN-600 in the first ten days of January 2010. The
observations were made using continuum radiometers of the first feed for
wavelengths of 1.38, 2.7, 3.9, 6.25, 13, and 31 cm. As in \cite{kh_a}, owing to
the poor interference conditions during the observations, data suitable for
the analysis were obtained in only four bands: 2.7, 3.9, 6.25, and 13 cm. At
the observed elevations, the beamwidths in the central cross section were
25", 36", 57", and 119", respectively. At 6.25 and 13 cm, we used subchannels
of the spectrum analyzer that enabled us to effectively
suppress interference. The observed sources are listed in Table 1, and a
log of the observations is in Table 2. Note that GRG\,2103+64 was observed
 nine times in this set of observations; however, we could not achieve a
signal-to-noise ratio suffcient to detect this source.

\begin{table*}
\begin{center}
\caption{Main parameters of the GRGs from the observing proposal}
\begin{tabular}{|ccccrr|}
\hline
Source        &Coordinates     &Redshift&Type&Angular size,&Flux density   \\
	      &RA+Dec (J2000.0)&        &    &    arcmin   &(1.4 GHz), mJy  \\
	      &hhmmss + ddmmss &        &    &             &                \\
\hline
GRG0452+5204 & 045253+520447   & 0.109   & I   & 9.7     & 2869.1  \\
GRG 0751+4231 & 075109+423124   & 0.203   & II  & 6.0     & 162.3   \\
GRG 0929+4146 & 092911+414646   & 0.365   & II  & 6.6     & 165.5   \\
GRG 1216+4159 & 121610+415927   & 0.243   & II  & 5.2     & 415.5   \\
GRG 1521+5105 & 152115+510501   & (0.37)  & II  & 4.3     & 1197.5  \\
GRG 1918+516  & 191923+514334   & 0.284   & II  & 7.3     & 920     \\
GRG 2103+6456 & 210314+645655   & 0.215   & II  & 4.8     & 119.7   \\
\hline
\end{tabular}
\end{center}
\end{table*}

\begin{table*}
\begin{center}
\caption{Observed GRG areas
(Central cross sections were done for all objects. $N_t$ is the number of
transits)}
\begin{tabular}{|ccr|}
\hline
Source       &center of the observed &  \\
	     &area RA+Dec (J2000.0)  &$N_t$   \\
	     &(hhmmss + ddmmss)      &  \\
\hline
GRG 0452+5204  & 045343.7+520556     &  11   \\
\hline
GRG 0751+4231  & 075153.9+422945     &  10   \\
\hline
GRG 0929+4146  & 092951.8+414353     &  10   \\
\hline
GRG 1216+4159  & 121641.4+415545     &  11   \\
\hline
GRG 1521+5105  & 152132.5+510232     &  7    \\
\hline
GRG 2103+6456  & 210322.1+645929    &  9     \\
\hline
\end{tabular}
\end{center}
\end{table*}

\subsection{Data Processing}

To reduce the flux densities to the international scale \cite{baars}, we observed
calibration sources from the standard RATAN-600 list \cite{aliakberov},\cite{trushkin}. The transit
curves of the sources were analyzed in the FADPS processing system
\cite{ver_c},\cite{ver_d}. We first subtracted a low-frequency trend with a smoothing window
of 8$\arcmin$ from the transit records of the sources. To estimate the flux densities,
we used the integrated values under the RATAN-600 beam transit curves
approximated by a set of Gaussians. The noise levels in the single-transit
records at 1.38, 2.7, 3.9, 6.25, and 13 cm at
an elevation of $82^\circ$ were 17.2, 8.9, 18.1, 10.7, and 96.6 mK/s1/2,
respectively. Table 3 lists our flux-density measurements at 2.7, 3.9, 6.25,
and 13 cm, together with the integrated flux densities for the measured
sources computed from the 21-cm maps of the NRAO VLA Sky Survey (NVSS, USA)
\cite{condon} and the 92-cm data of the Westerbork Northern Sky Survey (WENSS, the
Netherlands) \cite{rengelink}. Figure 1 presents the maps of the sources. To identify
the objects and estimate their parameters, we also used the CATS database
\cite{ver_e}.

\begin{table*}
\begin{center}
\caption{Integrated flux densities (in mJy), RATAN-600, WENSS, and NVSS data}
\begin{tabular}{|r|rrrrrrr|}
\hline
Source    &2.7 cm&3.9 cm&6.25 cm&13.5 cm&6.2 cm&21 cm&92 cm\\
name      &RATAN & RATAN& RATAN &RATAN  &RATAN &RATAN&RATAN \\
\hline
GRG0452+5204  & 417   &  827  & 1141   & 1984    & 844  & 3003 & 18760  \\
GRG 0751+4231  & 103   &  227  & 274    &  476    &  35  & 202  & 1365   \\
GRG 0929+4146  & --    &  --   & 215    &  315    &  91  & 200  & 1560   \\
GRG 1216+4159  & --    &  --   & 123    &  207    &  77  & 264  & 1604   \\
GRG 1521+5105  & --    &  --   & 317    &  549    & 377  & 747  & 4903   \\
GRG 2103+6456  & --    &  --   &$<$54   &$<$180   &  32  & 124  &  337   \\
\hline
\end{tabular}
\end{center}
\end{table*}

Among the CATS catalogs, we found identifications in the GB6 \cite{gregory}, VLSS \cite{cohen},
6C \cite{hales_a}, 7C \cite{riley}, 8C \cite{hales_b}, Texas \cite{douglas}, and B3 \cite{ficarra} surveys.

GRG\,0452+5204 and GRG\,0751+4231 were observed at 2.7 and 3.9 cm in a
beam-switching mode. To take into account the probable flux-density
decrease of an extended object in this observing mode, we modeled the transit
 of the sources across the two horns. Our modeling included calculation of
the two-dimensional RATAN-600 beam at the observed wavelength using the
method of Korzhavin \cite{korzhavin} and
the FADPS system \cite{ver_c}, convolution with the components of the observed
source, and calculation of a model transit across the RATAN-600 beam. The
corrections to the integrated flux densities of extended radio sources in
the modeled beam-switching mode were included in our analysis of the actual
observations.

The flux-density uncertainties for the RATAN-600 data were estimated from
the noise levels, and
were 2 mJy at 2.7 cm, 4 mJy at 3.9 cm, 5 mJy at 6.25 cm, and 52 mJy at 13 cm.

\subsection{Spectra}

Using our measurements, we plotted the the integrated radio spectra for the
six studied sources. We parametrized these spectra using the formula $ lg\,S(\nu) = A + Bx + C f(x)$,
where S is the flux density in Jy, x the logarithm of the frequency $\nu$ in
 MHz, and f(x) one of the following functions: exp(-x), exp x, or x2. We
used the "spg" system for the analysis of the spectra \cite{ver_e}. Figure 2 shows
the GRG spectra. Table 4 presents the analytical fits for the continuum
spectra of the studied GRGs.

\begin{table*}[!th]
\begin{center}
\caption{Fits for the GRG radio continuum spectra from 92 to 2.7 cm}
\begin{tabular}{|rr|}
\hline
Source name  &Radio spectrum \\
	     &               \\
\hline
GRG 0452+5204  &  $ 3.054-0.829x $ \\
GRG 0751+4231  &  $ 1.971-0.697x $ \\
GRG 0929+4146  &  $ 1.330-0.583x $ \\
GRG 1216+4159  &  $ 2.288-0.882x $ \\
GRG 1521+5105  &  $ 2.233-0.731x $ \\
GRG 2103+6456  &  $ 1.942-0.923x $ \\
\hline
\end{tabular}
\end{center}
\end{table*}

\section{DISCUSSION}

We have observed five GRGs for the first time and measured their flux
densities using the RATAN-600 radiometer complex. We have calculated
the spectral indices of the GRGs as the slopes of the tangents to the
approximating curve for the integrated spectrum at a given frequency.
Table 3 lists the flux densities measured at 2.7, 3.9, 6.25, and 13 cm,
and Table 4 the spectral indices, equal to the coeffcients of the argument
x. The RATAN-600 data for GRG\,0751+4231 lie above the GB6 and NVSS data.
The object from the GB6 catalog is virtually pointlike; this explains the
low level of its catalog flux. It is most likely that the level of the NVSS
flux density has the same origin, since integrating over the peak
values results in the incomplete detection of weak diffuse emission.

The radio source observed on RATAN-600 in the area of GRG\,0929+4146 consists
of two radio galaxies: GRG\,0929+4146 proper, as a multi-component object
extended along a line, and a FR II double
radio source with coordinates $\alpha$ = 09h29m24s, $\delta$ =
+41$^\circ$46'18" that merges into a single extended object at the left,
near GRG\,0929+4146 (Fig. 1). The
RATAN-600 does not resolve these radio galaxies in a meridian transit; thus,
Figure 2 presents the total spectrum. Using the NVSS, WENSS, and 7C data,
we have plotted separate spectra for each of the radio galaxies (Fig. 3).
The integrated radio spectrum for GRG\,0929+4146 alone is approximated as
y = 3.046 - 1.208x, and the spectrum of the neighboring radio galaxy is
y = 1.818 - 0.774x, with a smaller slope, demonstrating that the RATAN-600
observations at short wavelengths are dominated by the radio emission from
J092924+414618.

\begin{figure*}
\centerline{
\vbox{
\hbox{
  \psfig{figure=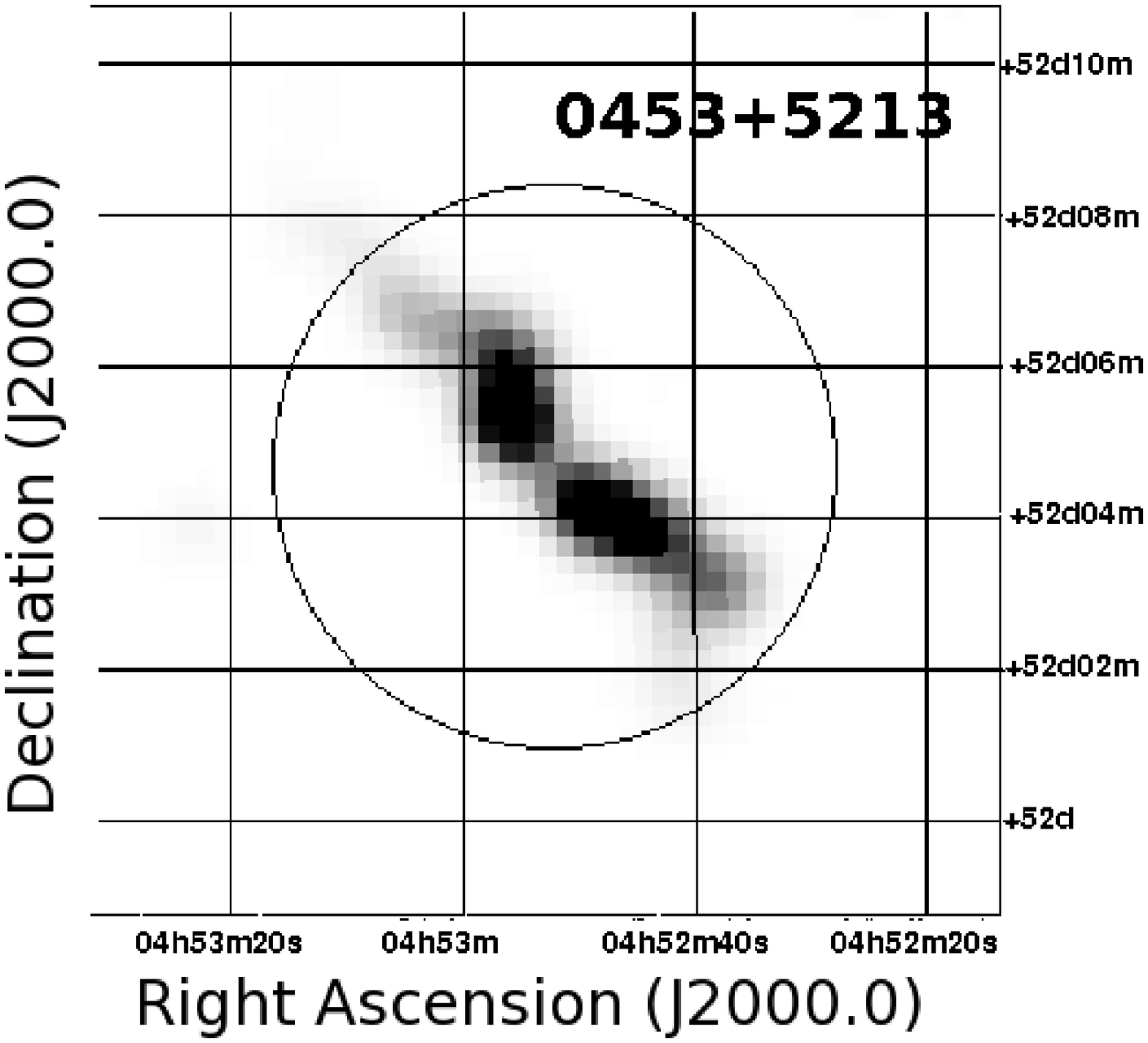,width=7.5cm,angle=-90}
  \psfig{figure=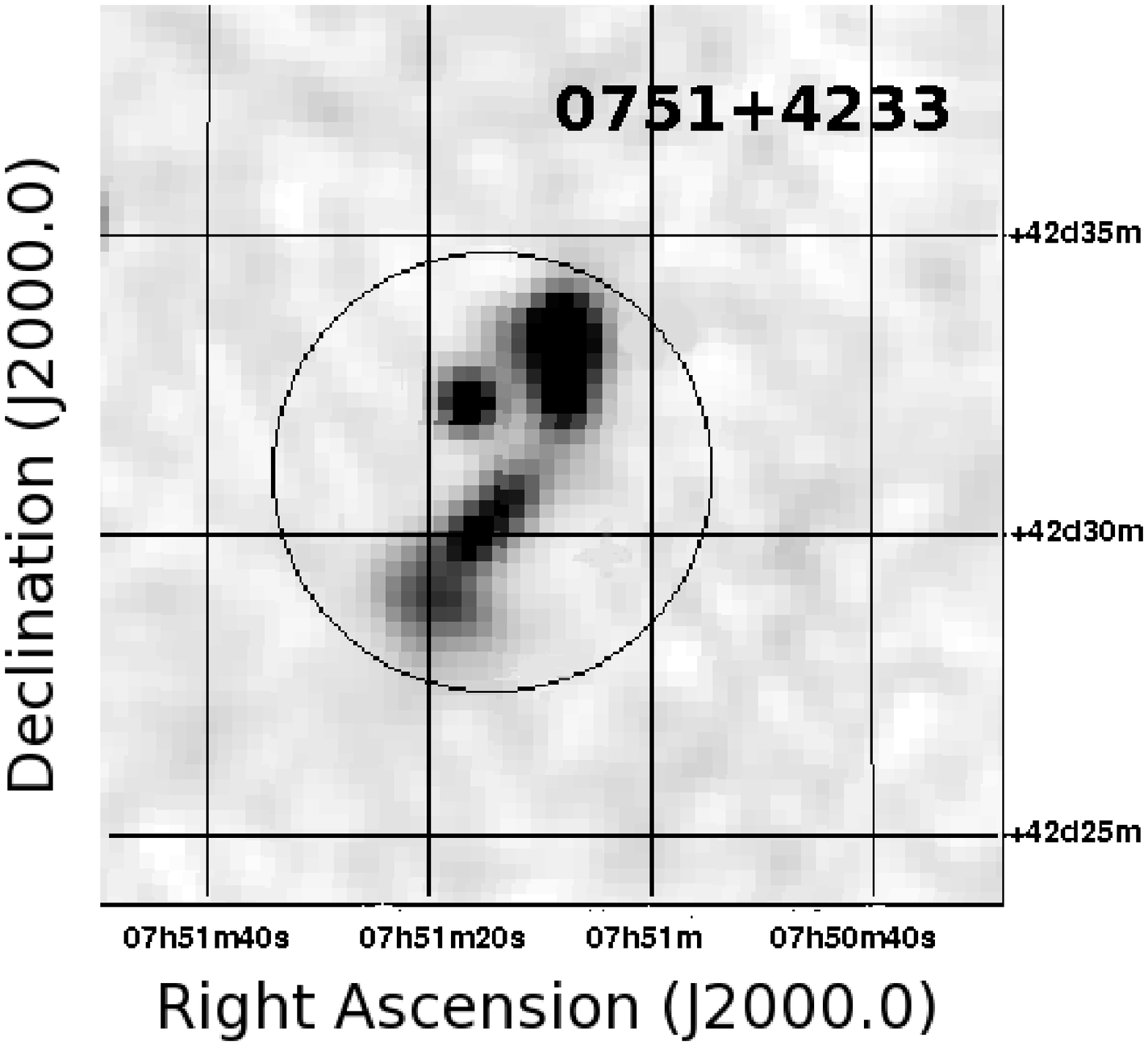,width=7.5cm,angle=-90}
}
\hbox{
  \psfig{figure=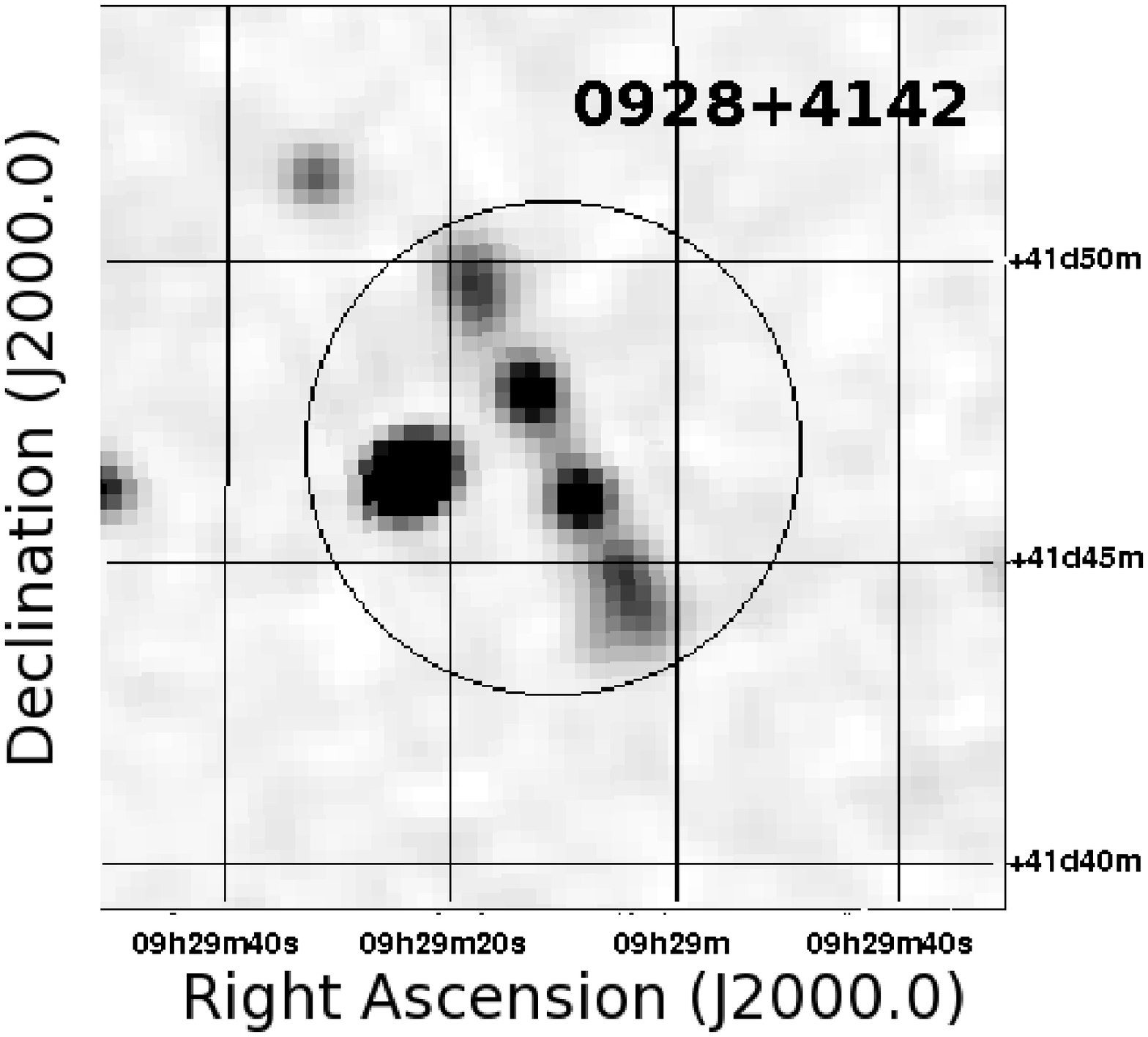,width=7.5cm,angle=-90}
  \psfig{figure=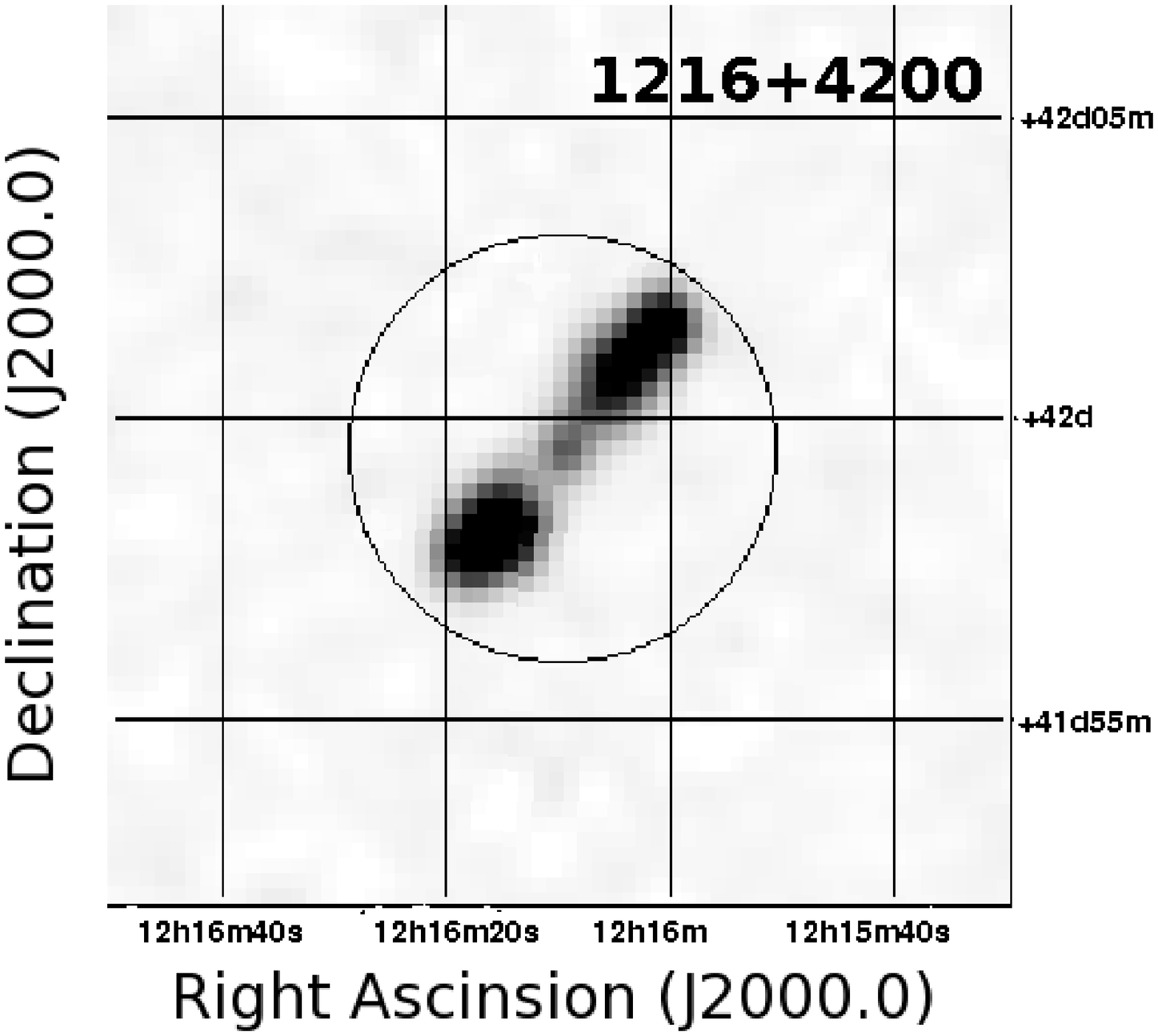,width=7.5cm,angle=-90}
}
\hbox{
  \psfig{figure=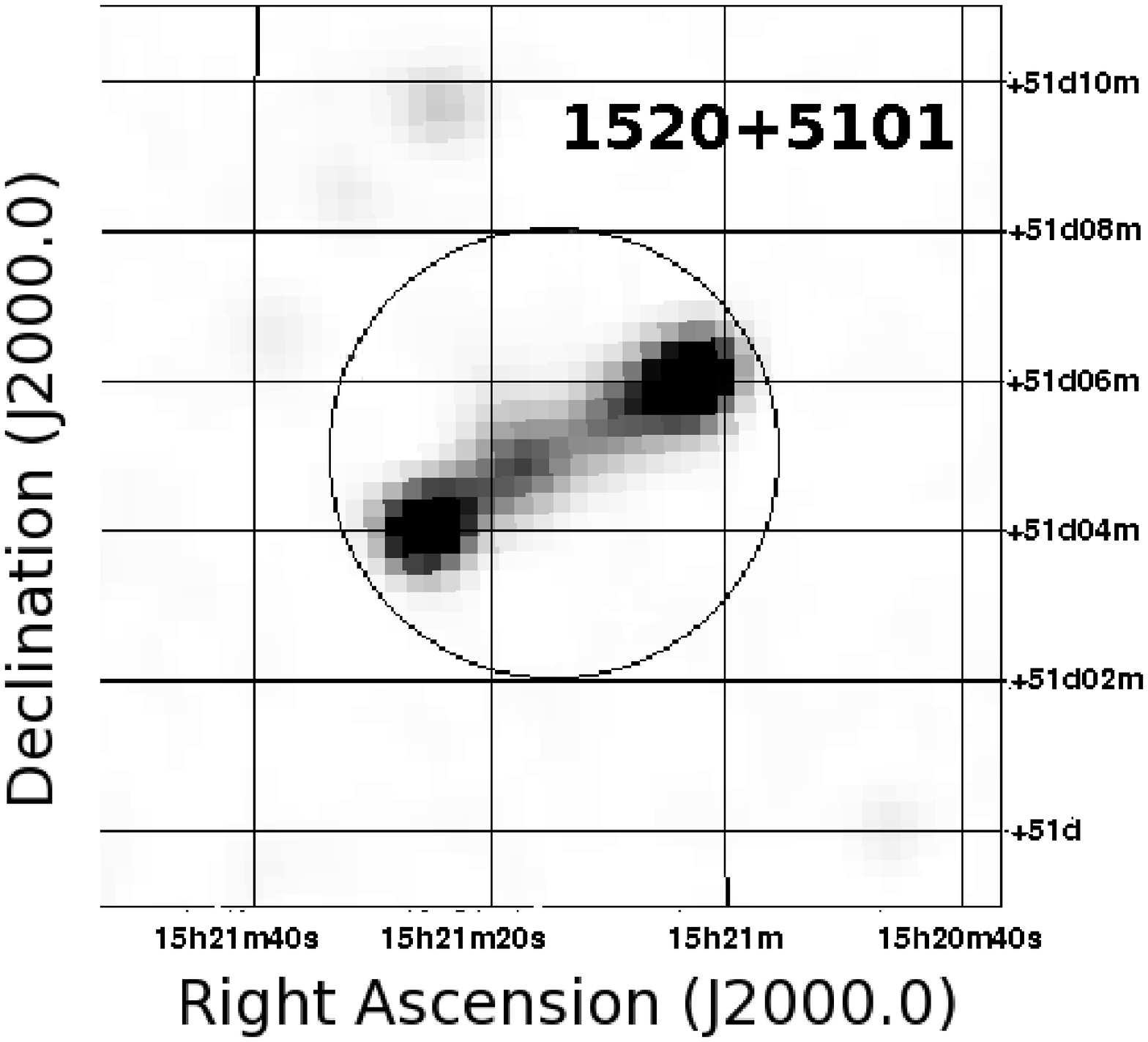,width=7.5cm,angle=-90}
  \psfig{figure=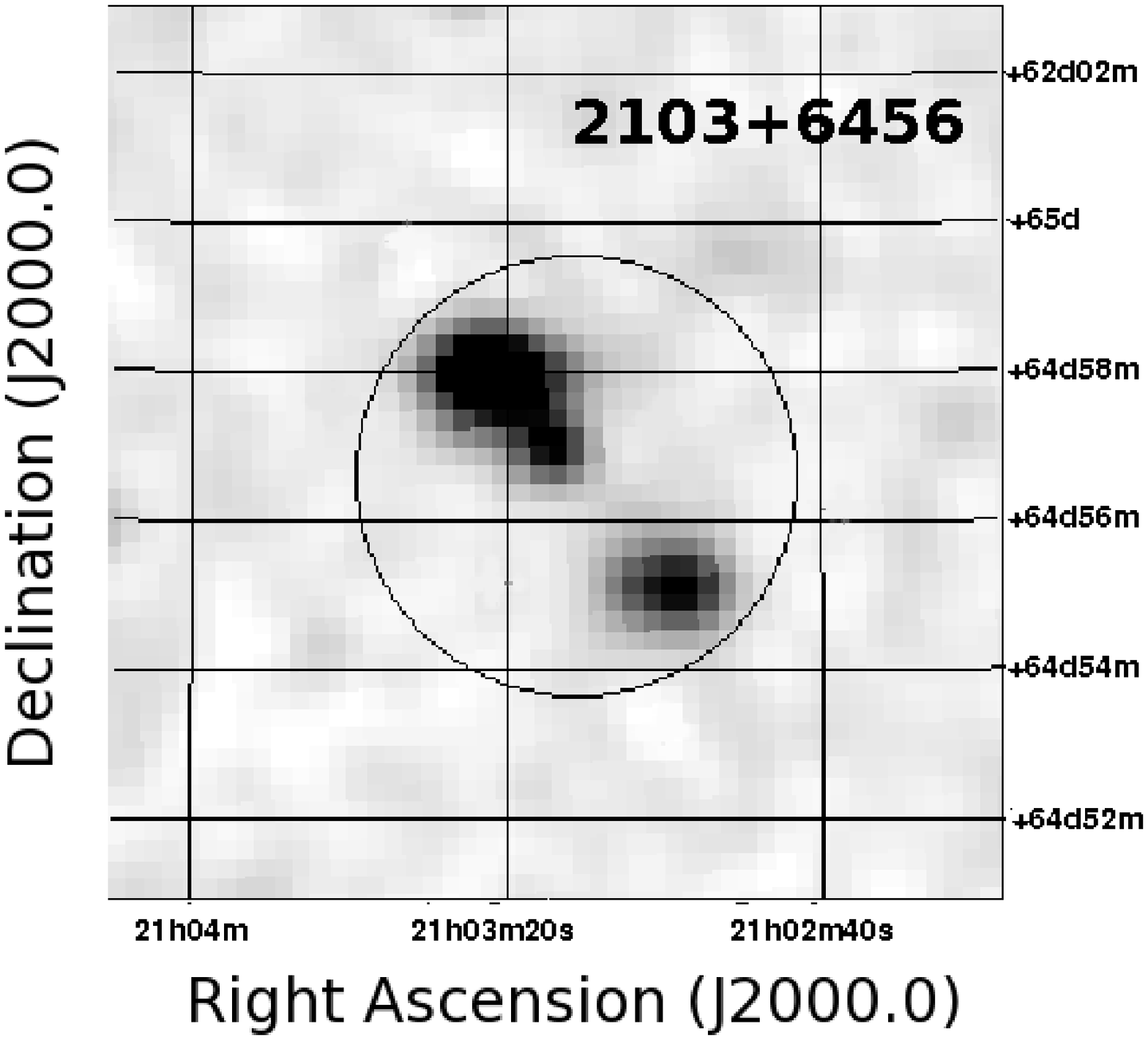,width=7.5cm,angle=-90}
}
}}
\caption{NVSS radio images of GRGs. The circles denote the objects observed
on the RATAN-600.}
\end{figure*}

\begin{figure*}
\centerline{
\vbox{
\hbox{
\psfig{figure=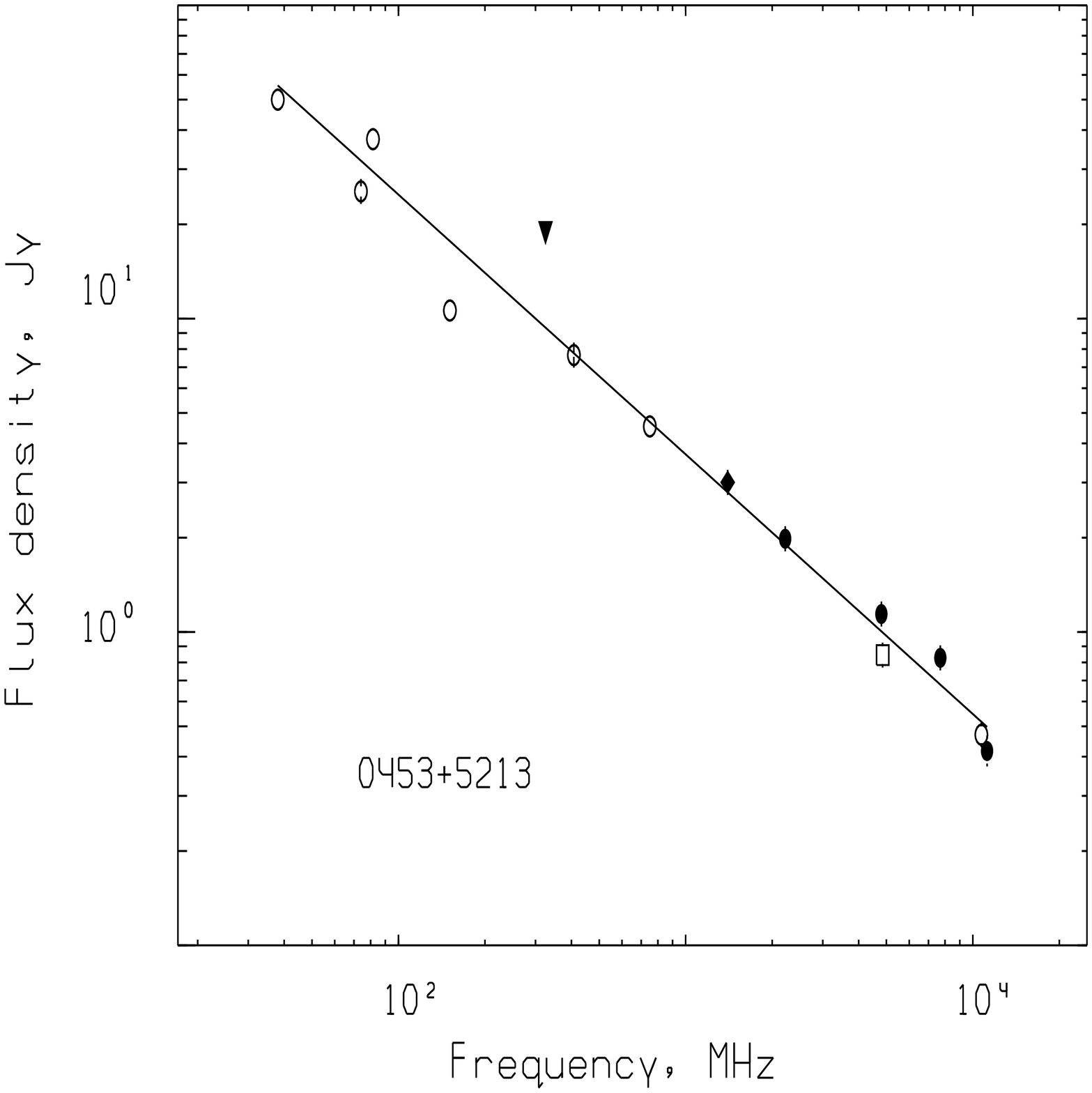,width=5cm}
\mbox{\hspace*{2cm}}
\psfig{figure=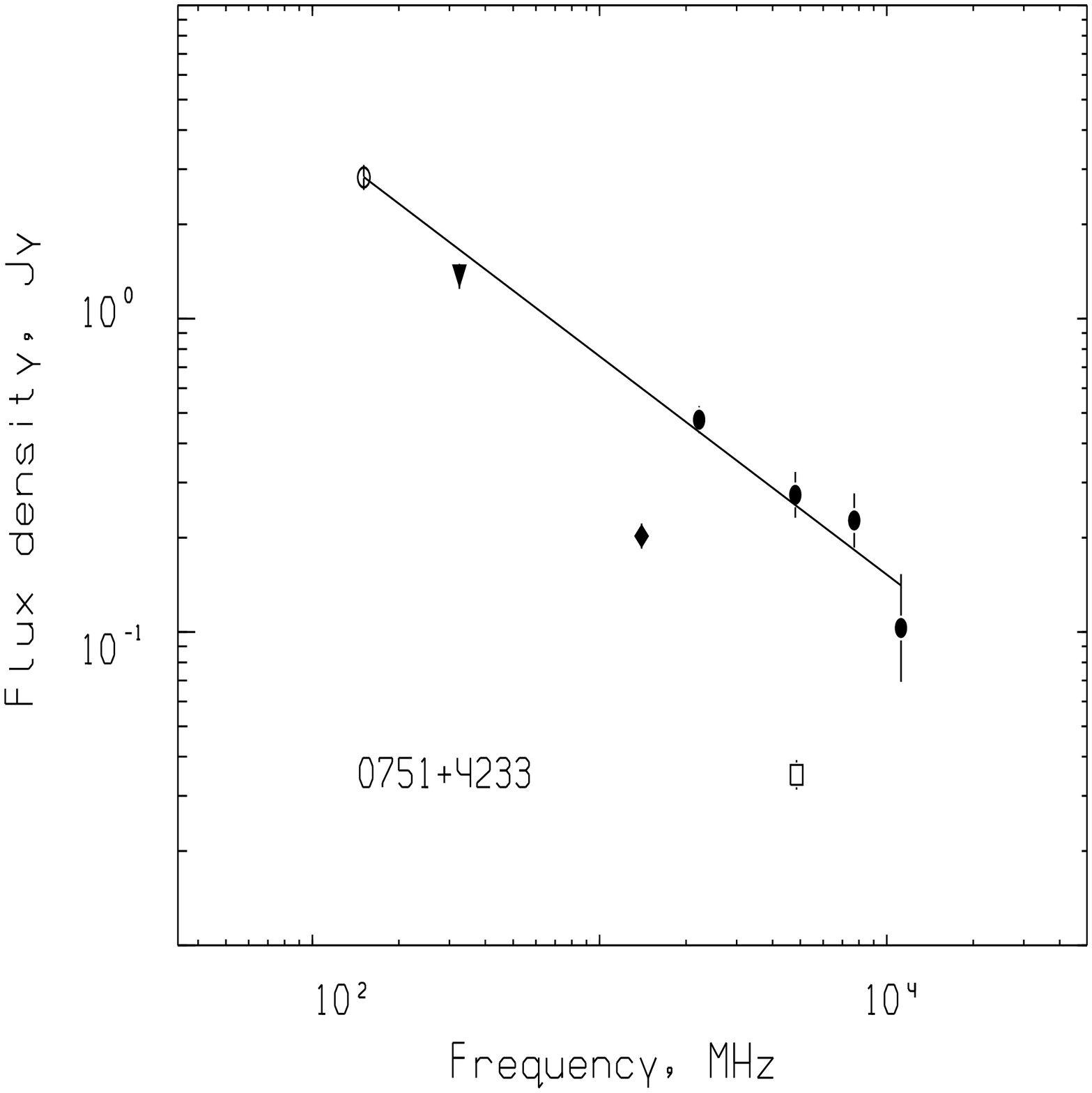,width=5cm}
}
\hbox{
\psfig{figure=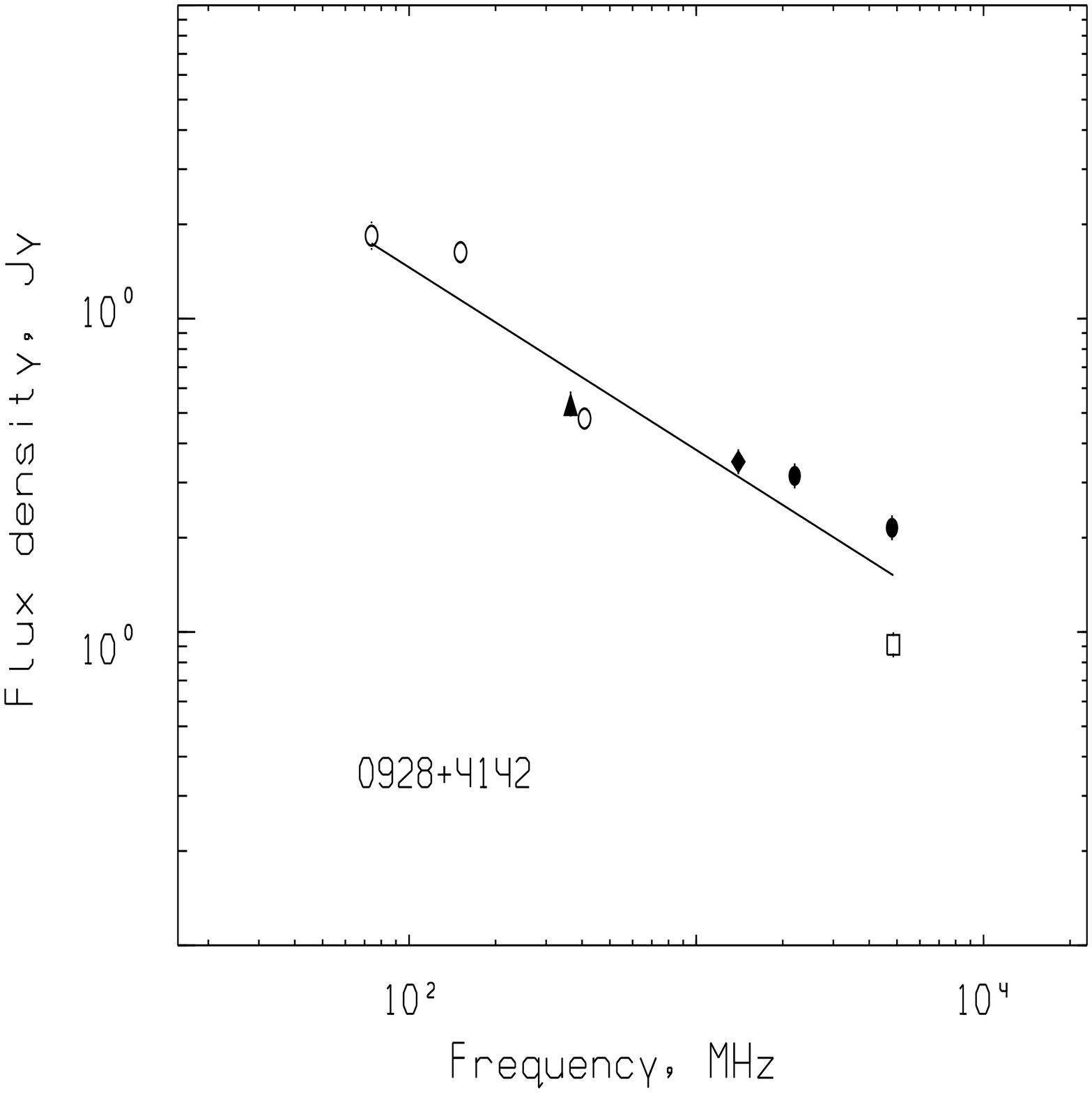,width=5cm}
\mbox{\hspace*{2cm}}
\psfig{figure=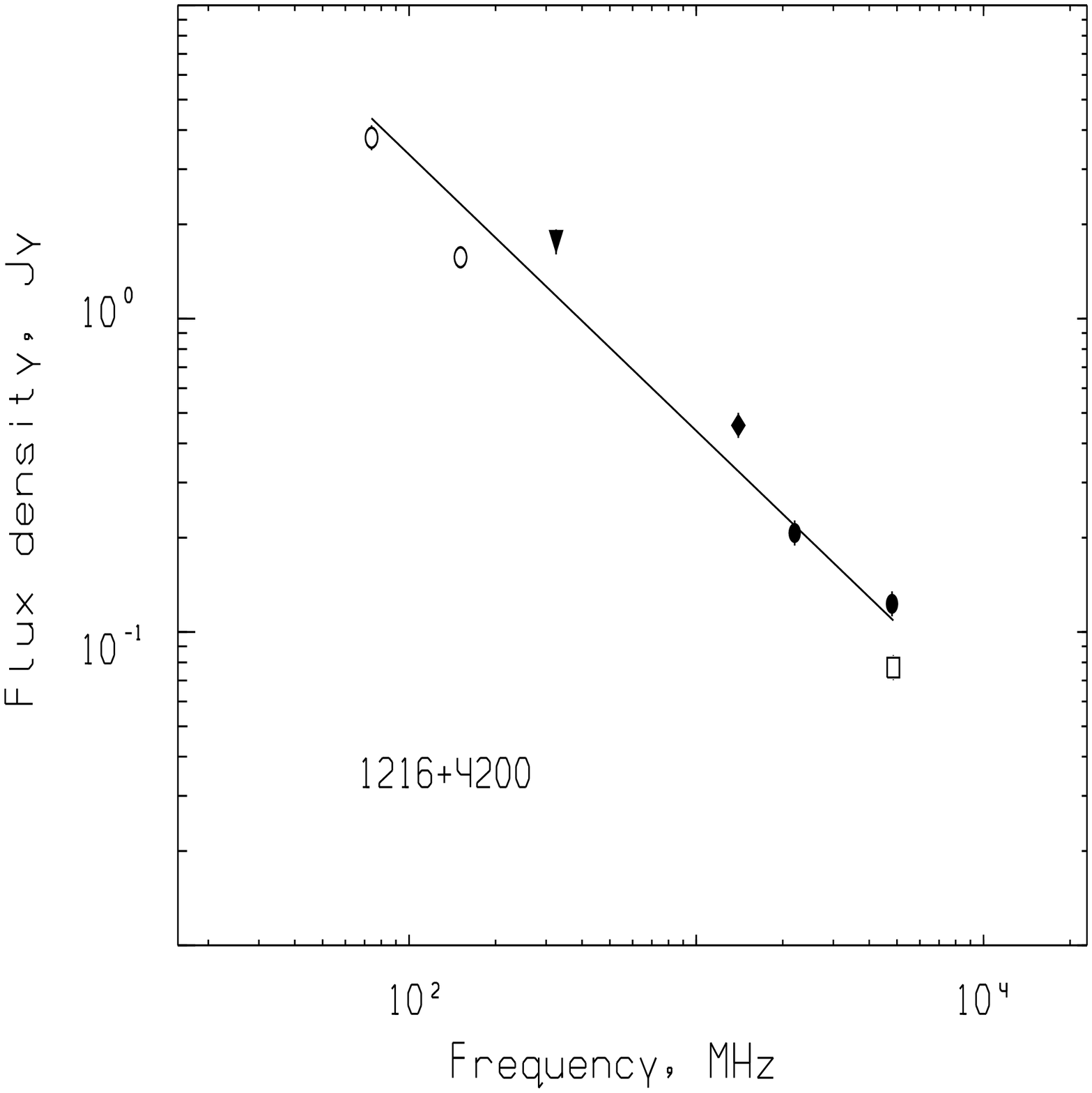,width=5cm}
}
\hbox{
\psfig{figure=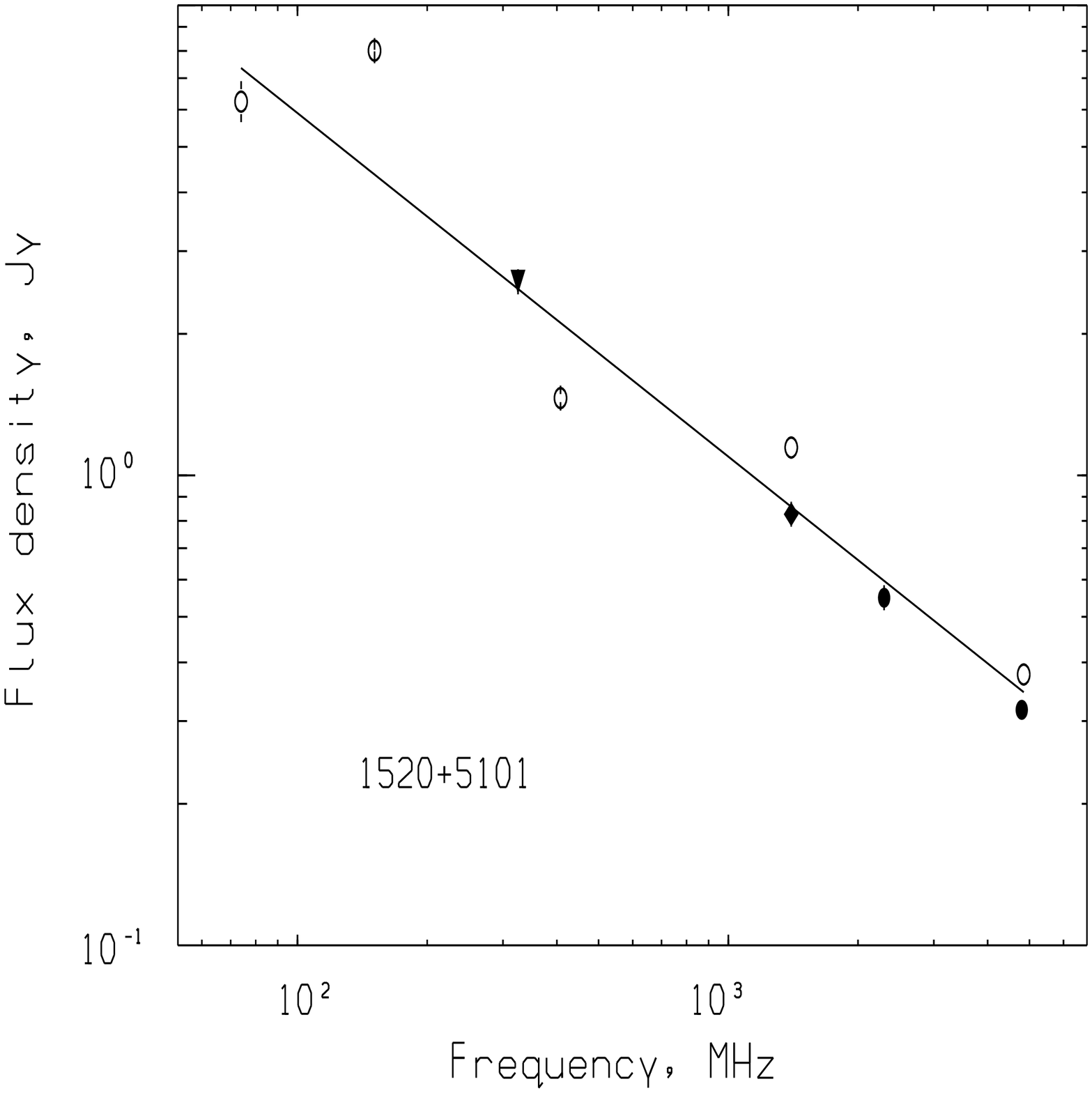,width=5cm}
\mbox{\hspace*{2cm}}
\psfig{figure=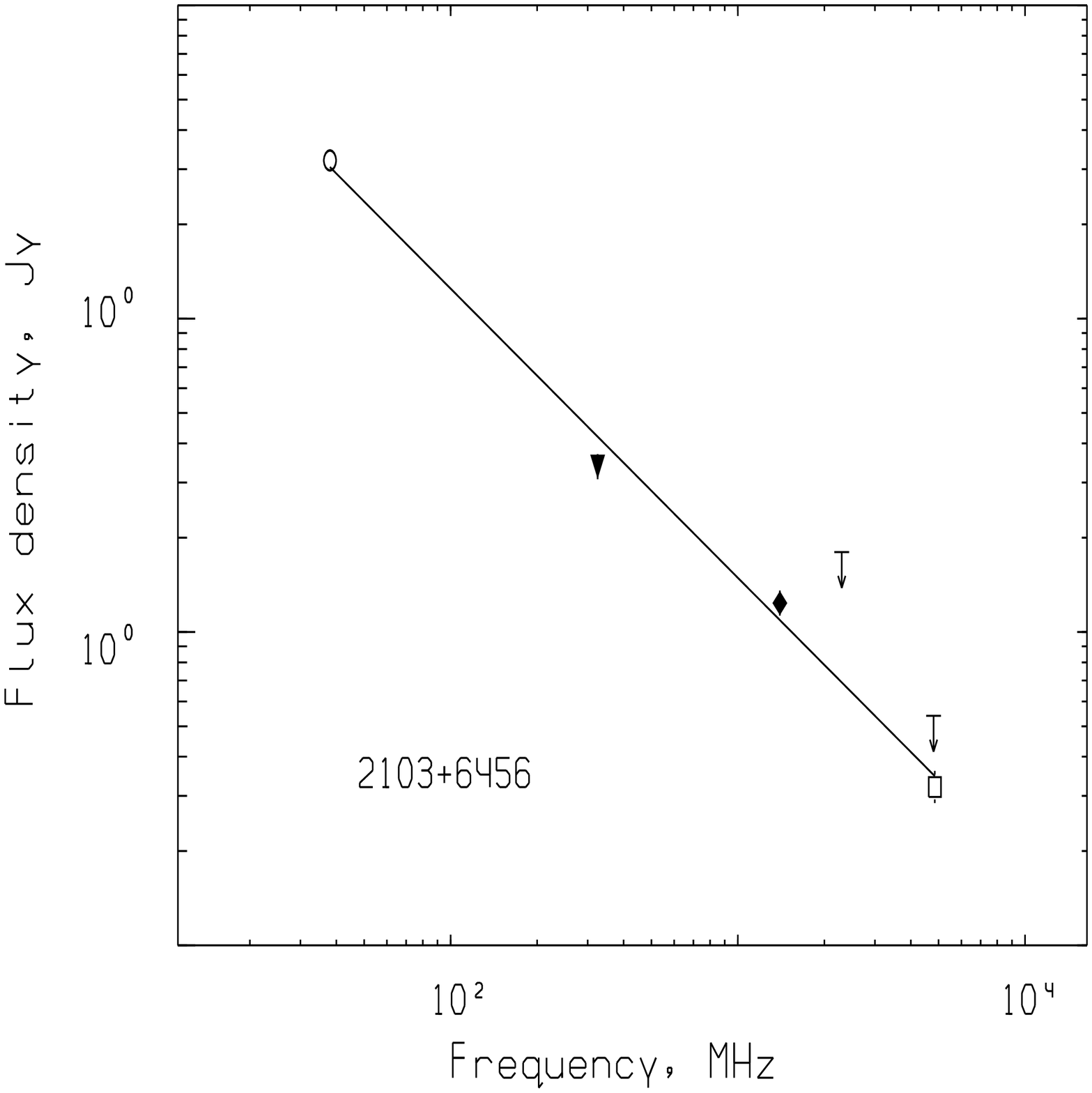,width=5cm}
}
}}
\caption{Radio spectra of the GRGs plotted with data from the RATAN-600,
NVSS, WENSS, GB6 (Table 3), etc. observations. The RATAN-600 data are shown
by filled circles.}
\end{figure*}

\begin{figure}
\centerline{\psfig{figure=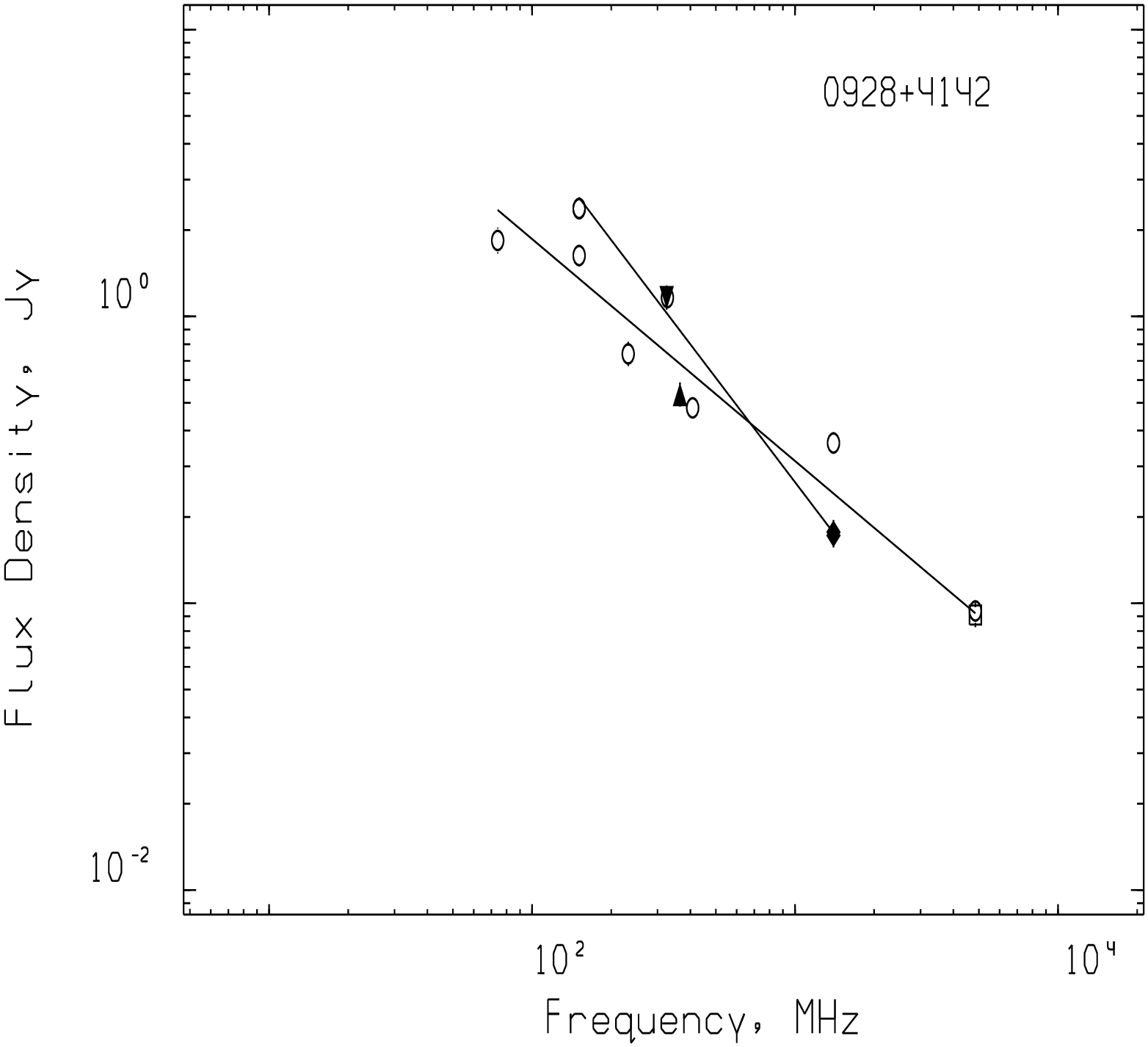,width=6cm}}
\caption{
Integrated radio spectra of the radio galaxies (RATAN-600 observations):
GRG 0929+4146, which has a steeper spectrum, and the ordinary double radio
galaxy J092924+414618.}
\end{figure}

We resolved GRG\,1521+5105 into two components, whose spectra are presented
in Fig. 4. The integrated flux densities for one component (J152103+510600)
were 368 mJy at 13 cm and 167 mJy at 6.25 cm, whereas they were 181 mJy at
13 cm and 150 mJy at 6.25 cm for the other
component (J152125+510401). The functions fitting the radio spectra of the
components are y = 2.226 - 0.800x and y = 1.537 - 0.645x, respectively.
GRG\,1521+5105, which is identified with the galaxy SDSS J152114.55+510500.9
and has the photometric redshift z = 0.37 (NED data\footnote{\tt http://ned.ipac.caltech.edu/}),
is projected against
the outskirts of the cluster NSCS J152018+505306, with redshift z = 0.52
(NED), at an angular distance of 15' from its center.
However, there are more than 1700 galaxies (NED data) and a great number of
radio sources within 10' of the radio galaxy (Fig. 4). Due to the lack of
redshifts for the overwhelming majority of these, it is di cult to judge
whether there could be a physical
association of GRG\,1521+5105 with any group of galaxies. However, the rich
environment of this radio galaxy stimulates additional interest in searching
for the origin of its giant size.

\begin{figure*}
\centerline{
\hbox{
\psfig{figure=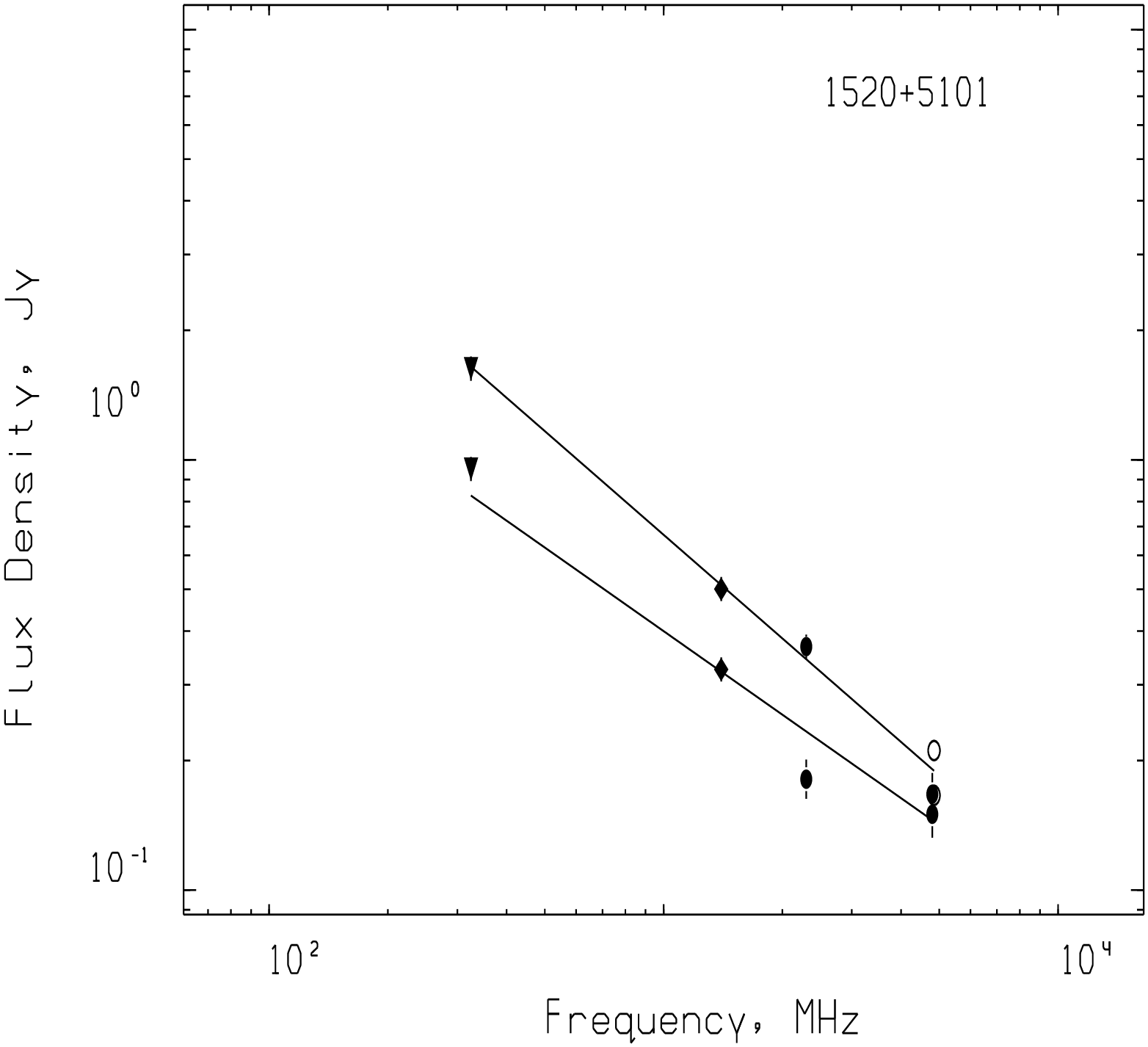,width=5cm}
\mbox{\hspace*{2cm}}
\vbox{
\psfig{figure=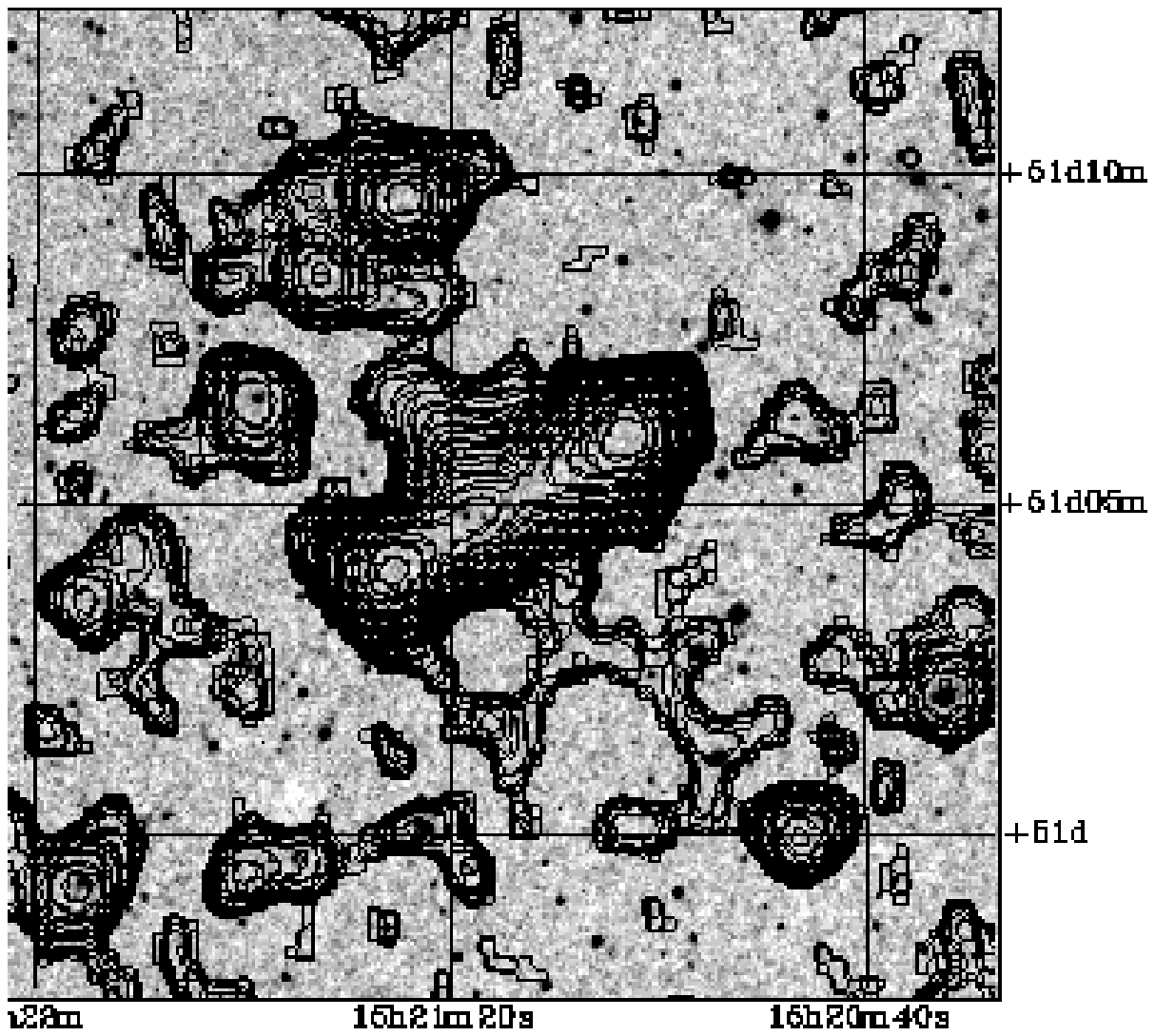,width=6cm}
}
}}
\caption{
Left: radio spectra of the components of GRG\,1521+5105, with the RATAN-600
data shown as filled circles. Right:
NVSS radio image of GRG\,1521+5105 (at the map center), overlaid on the
DSS optical image.}
\end{figure*}

In addition, as in our previous work \cite{kh_a}, we have estimated the contribution
of the observed GRGs to the microwave background. We used the RATAN-600 data
and the updated GRG spectra to calculate the integrated flux densities at
millimeter wavelengths, listed in Table 5. The contribution of the studied
giant radio sources on scales of galaxy clusters exceeds 1 mJy, and could
be an important confusion factor in distinguishing components of the
microwave background.

We envisage further accumulation of new data, including compiling lists of
new GRGs and observations on the RATAN-600.

\section{ACKNOWLEDGMENTS}

The authors thank Yu.V.~Sotnikova for help with the RATAN-600 observations.
This research has made use of the NASA/IPAC Extragalactic Database (NED),
which is operated by the Jet Propulsion Laboratory, California Institute of
Technology, under contract with the National Aeronautics and Space
Administration. We have also used the CATS database\footnote{\tt http://www.sao.ru/cats/}
\cite{ver_e} and FADPS
system for processing radio astronomy data\footnote{\tt http://sed.sao.ru/~vo/fadps\_e.html}
\cite{ver_c},\cite{ver_d}. We are deeply grateful
to R.D. Dagkesamanskii for valuable comments on the manuscript. This work
was supported by the Program of State Support of Leading Scientific Schools
of the Russian Federation, the Russian Foundation for Basic Research
(project 09-02-92659-IND), and the Dynasty Foundation of Nonprofit Programs.

\end{document}